\documentclass{eas}
\usepackage{graphicx}
\begin{document}

\title{SPICA: the next generation\\Infrared Space Telescope} 
\runningtitle{SPICA: the next generation IR Space Telescope}
\author{Javier R. Goicoechea}\address{Departamento de Astrof\'{\i}sica. Centro de Astrobiolog\'{\i}a (CSIC-INTA), 
Madrid, Spain\\$^2$Institute of Space and Astronautical Science (ISAS)
and Japan Aerospace Exploration Agency (JAXA). Yoshinodai, Sagamihara, Kanagawa 229-8510, Japan.}
\author{Takao Nakagawa$^2$ on behalf of the SAFARI/SPICA teams} 
\begin{abstract}
We present an overview of SPICA, 
the \textit{Space Infrared Telescope for Cosmology and Astrophysics},
a world-class space observatory optimized for mid- and far-IR
astronomy (from 5 to $\sim$210$\mu$m) with  a cryogenically cooled $\sim$3.2\,m telescope ($<$6\,K). 
Its high spatial resolution and  unprecedented sensitivity in both photometry and
spectroscopy modes will enable us to address a number of key problems in 
astronomy.
SPICA's large, cold aperture will provide a two order of magnitude  
sensitivity  advantage over current far--IR facilities 
($\lambda$$>$30\,$\mu$m wavelength). 
In the present design, SPICA will carry mid-IR camera, spectrometers 
and  coronagraph (by JAXA institutes) and a far-IR imager FTS-spectrometer, SAFARI 
($\sim$34-210\,$\mu$m, provided by an European/Canadian consortium lead by SRON). 
Complementary instruments such as  a far-IR/submm spectrometer 
(proposed by NASA) are also being discussed.
SPICA will be the only space observatory of its era to bridge the far--IR wavelength gap 
between JWST and ALMA, and carry out unique science not achievable at visible or
submm wavelengths.
In this contribution we summarize some of the  scientific advances 
that will be made possible  by the large increase in sensitivity compared to previous infrared space 
missions.
\end{abstract}
\maketitle
\section{Introduction}

Understanding of the origin and evolution of galaxies, stars, planets, our Earth and of 
life itself are fundamental objectives of Science in general and Astronomy in particular. 
Although impressive advances have been made in the last twenty years, our knowledge of how 
the Universe has come to look as it does today is far from complete. 
A full insight of the processes involved is only possible with observations in the long wavelength
infrared waveband of the electromagnetic spectrum. 
It is in this range that astronomical objects emit most of
their radiation as they form and evolve in regions where obscuration by dust prevents 
observations in the visible and near infrared.
Over the past quarter of a century successive space infrared observatories 
(IRAS, ISO, Spitzer and AKARI) have revolutionized our understanding of the evolution 
of stars and galaxies. Mid to far infrared observations have led to stunning discoveries such as 
the Ultra Luminous Infrared Galaxies (ULIRGS), the basic processes
of star formation from ``class 0'' protostars through to the clearing of the gaseous 
proto-planetary disks and the presence of dust excesses around main sequence stars. 
The Herschel Space Observatory launched in 2009 continues this work in the far infrared and sub-mm and JWST, 
due for launch around 2014, 
will provide a major boost in observing capability in the 2--28\,$\mu$m range.
Previous infrared missions have been hampered by the requirement to cool the telescope and 
instruments to $<$5~K using liquid cryogens. This has limited the size of the apertures to $<$1\,m 
and our view of the infrared Universe has been one of poor spatial resolution and limited 
sensitivity. The \textit{Herschel} mission addresses the first of these by employing a 
3.5\,m mirror to dramatically increase the available spatial resolution but, because
it is only cooled to $\sim$80\,K (Pilbratt et al. 2010), only offers a modest increase in sensitivity in the 55--210$\mu$m 
range compared to previous facilities. JWST will provide a major increase in both spatial 
resolution and sensitivity but only up to 28$\mu$m.

\section{SPICA}

The JAXA led mission SPICA is an observatory that will provide imaging and spectroscopic capabilities in
the 5 to 210\,$\mu$m wavelength range with a $\sim$3.2\,m telescope (current design), 
but now cooled to a temperature of $\sim$6 K. 
In combination with a new generation of highly sensitive detectors, the low telescope temperature
will allow us to achieve sky-limited sensitivity over the full 5 to 210$\mu$m band for the first time. This unique
capability means that SPICA will be between one and two orders of magnitude more sensitive than Herschel
in the far infrared band. SPICA will cover the full 5 to 210\,$\mu$m wavelength range, including the missing 
28-55$\mu$m octave which is out of the Herschel and JWST domains, with unprecedented sensitivity and spatial
resolution. Furthermore, SPICA will be the only observatory of its era to bridge the wavelength gap between
JWST and ALMA, and carry out unique science not achievable at (sub)mm wavelengths with ALMA. In the
mid infrared SPICA will be able to carry out medium and high-resolution ($R\sim30,000$) spectroscopy, one order
of magnitude higher than in JWST, and will include spatial high multiplexing imaging and medium spectral
resolution capabilities. In addition, the characteristics of the SPICA monolithic telescope will provide unique
and optimal conditions for mid infrared coronagraphy in imaging and, uniquely, spectroscopic mode.

\subsection{Scientific Instruments}

At present, the scientific instruments foreseen for SPICA are:\\
$\bullet$ A mid infrared camera and spectrophotometer, MIRACLE (5-38$\mu$m, 6$'$$\times$6$'$ FOV).\\
$\bullet$ A mid infrared medium resolution spectrometer, MIRMES (10.32-36.04$\mu$m with
a spectral resolution of $R\sim$900-1500).\\
$\bullet$ A mid infrared high resolution spectrometer, MIRHES (4-18$\mu$m with a spectral
resolution of $R\sim$20,000-30,000).\\
$\bullet$ A mid infrared coronagraph, SCI (5(3.5)--27$\mu$m, $R\sim$20-200, contrast 10$^{-6}$, inner working angle 
$\sim$3.3$\lambda$/D$_{tel}$).\\
$\bullet$ SAFARI, a far infrared  imaging Fourier-Transform Spectrometer designed 
to provide continuous coverage in photometry and spectroscopy from 34 to 210\,$\mu$m, with a field of view of 2$'$$\times$2$'$ 
and spectral resolution modes of $R$$\sim$2000 (at 100$\mu$m), $R$ a few hundred and 20$<R<$50. The spectral sensitivity 
is required to be $\sim$3.5$\times$10$^{-19}$\,W\,m$^{-2}$ at 48\,$\mu$m (5$\sigma$, 1 hour). 
TES bolometer detectors will be used. SAFARI is to be built by a consortium of European institutes lead by SRON
(with Canadian and Japanese participation).

\begin{figure}[h]
  \centering 
  \includegraphics[height=0.48\hsize{},angle=0]{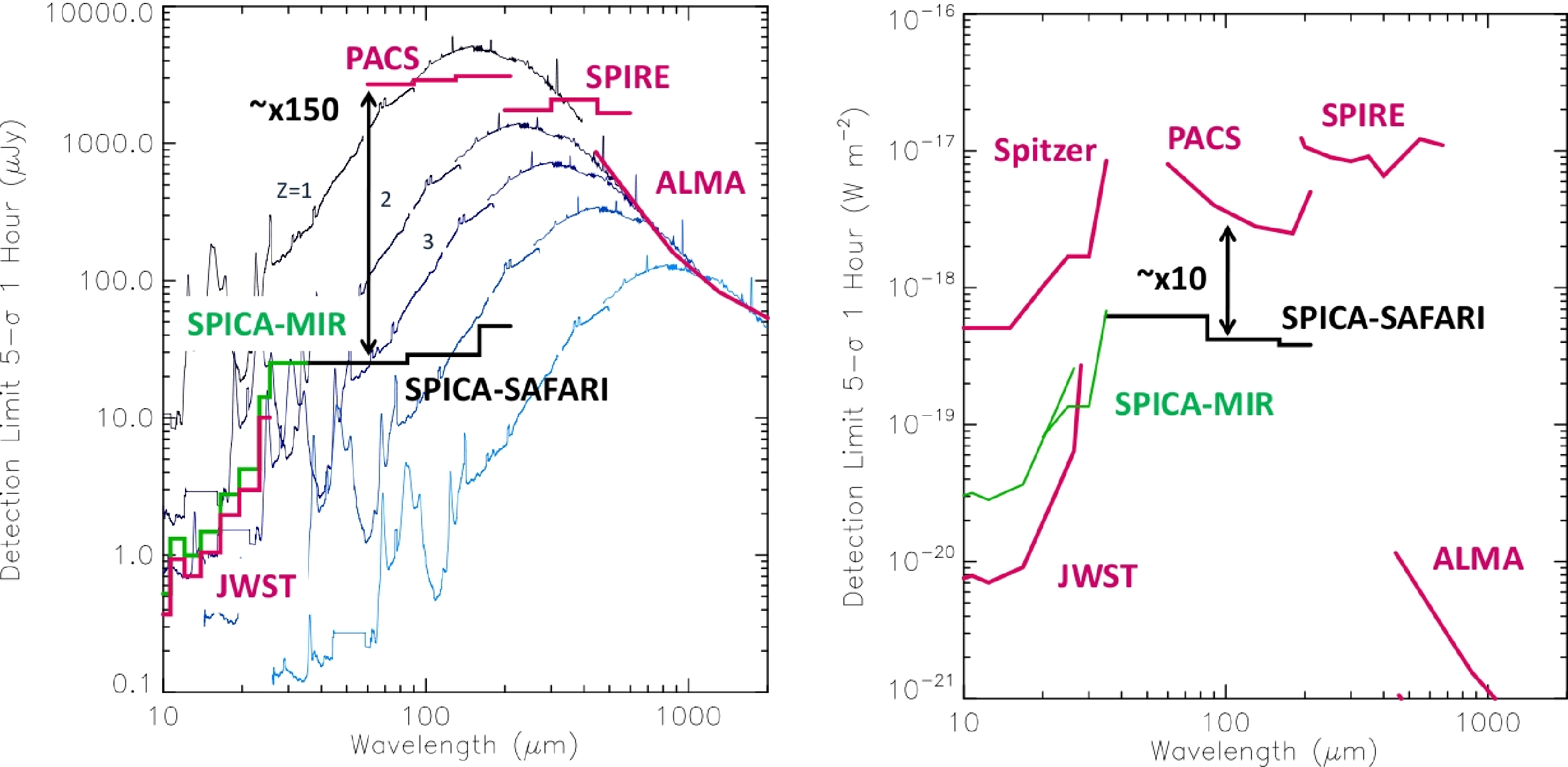} %
   \caption{\textit{Left panel}: Photometric performance expected for SPICA (green and black), 
compared to Herschel, ALMA and JWST (purple), for a point source (in $\mu$Jy for 5$\sigma$--1\,h). 
Note the $\sim$2 orders of magnitude increase in far-IR photometric 
sensitivity compared to Herschel-PACS.  For illustrative purposes the SED of the starburst galaxy
M82 as redshifted to the values indicated is shown in the background. \textit{Right panel}: Spectroscopic 
performance expected for SPICA (green and black) compared to predecessor and complementary 
facilities (purple) for an 
unresolved line for a point source (in~Wm$^{-2}$~for 5$\sigma$--1\,h). 
For ALMA, 100\,km\,s$^{-1}$ resolution is assumed. The SPICA MIR sensitivities are scaled by
telescope area from the JWST and Spitzer-IRS values respectively.
SPICA/SAFARI  sensitivity figures adapted from Swinyard, Nakagawa et al. (2009).}
  \label{fig:millenium}
\end{figure}

\subsection{Science with SPICA}

SPICA offers a sensitivity up to two orders of magnitude better than Herschel covering the mid-to-far-IR 
(mostly unreachable from the ground) with imaging, spectroscopic and coronagraphic
instruments (see Figure~1). Thanks to this tremendous increase in sensitivity, SPICA will make photometric
images in a few seconds that would take hours for Herschel and will take a full 5-210\,$\mu$m infrared spectrum
of an object in one hour that would take several thousand hours for Herschel. We illustrate this major increase
in sensitivity in Figure~2 which shows the area covered by a full spatial and spectral survey in
which we can detect spectroscopically all galaxies down to a luminosity of $\sim$10$^{11}$L$_{\odot}$ at z=1 and 
$\sim$10$^{12}$\,$L_{\odot}$ at z=2 in 900 hours. In theory, in about twice this time, the Herschel-PACS spectrometer 
would just be able to detect a single object over its full waveband to the same sensitivity. 
We can immediately see that this major increase in sensitivity, combined with a wide field of view and coverage 
of the full 5--210$\mu$m waveband, will revolutionize our ability to spectroscopically explore the nature of the 
thousands of objects that Herschel, JWST, and SPICA will discover in photometric surveys.\\

With its powerful scientific capabilities, SPICA will provide unique and ground-breaking answers to 
many key questions, and it is with this goal that the SPICA science objectives are being defined:

\textbf{(i) The formation and evolution of planetary systems}: By accessing key spectral diagnostic lines, SPICA
will provide a robust and multidisciplinary approach to determine the conditions for planetary systems formation.
This will include the detection for the first time of the most relevant species and mineral components in
the gas and dust of hundreds of transitional protoplanetary disks at the time when planets form. SPICA will also
be able to trace the warm gas in the inner ($<$30\,AU) disk regions and by resolving the gas Keplerian rotation
(\textit{e.g.,} by resolving the mid-IR H$_2$ lines),
will allow us to observe the evolution of disk structure due to planet formation.
SPICA will study debris disks and make the first unbiased survey of the presence of zodiacal clouds in
thousands of exoplanetary systems around all stellar types. It will allow us to detect both the dust continuum
emission and the brightest grain/ice bands as well as the brightest lines from any gas residual present in the
disk. SPICA will have the unique capability to observe water ice in all environments, and thus fully explore its
impact on planetary formation and evolution and in the emergence of habitable planets. In the closest debris
disks, SPICA will spatially resolve the distribution of water ice and determine the position of the ``snow line''
around Vega-like stars,
which separates the inner disk region of terrestrial planet formation from that of the outer planets.
SPICA will also drastically enhance our knowledge of the Solar System by making the first detailed characterization
of hundreds of Kuiper Belt Objects, and of different families of inner, hotter centaurs, comets and
asteroids. SPICA will provide the means to quantify their composition and determine unambiguously their size
distribution: critical observational evidence for the models of Solar System formation. No other planned or
present facility will be able to carry out these observations.
SPICA will provide direct imaging and low resolution mid infrared spectroscopy of outer young giant
exoplanets (\textit{e.g.,} at $\sim$10\,AU of a star at $\sim$10\,pc), which will allow us for the first time to study 
the physics and composition of their atmospheres in a wavelength range particularly rich in spectral signatures 
(\textit{e.g.,}~H$_2$O, CH$_4$, O$_3$, silicate clouds, NH$_3$, CO$_2$) and to compare it with
Solar System planet atmospheres. In addition, mid infrared transit photometry and  high resolution spectroscopy of 
``hot Jupiters'' will be routine with SPICA. Finally, by mapping very extended but faint regions,
SPICA will also provide an unprecedentedly sensitive window into 
key aspects of the dust life-cycle both in the Milky Way and in nearby galaxies, from its formation in evolved stars,
its evolution in the ISM, its processing in supernova-generated shock waves and massive stars, to its final 
incorporation into star forming cores and protoplanetary disks.

\begin{figure}[ht]
  \centering 
  \includegraphics[height=0.5\hsize{},angle=0]{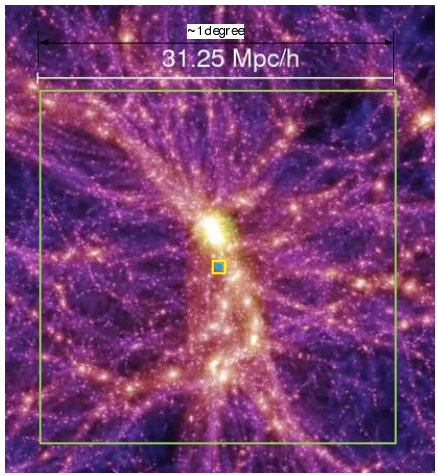} %
   \caption{Spectroscopic mapping speeds of the SPICA far-IR instrument (SAFARI) and Herschel-
PACS superposed on a realization of the Millenium simulation at z$\sim$1.4 (Springel et al. 2006).
In the center are the footprints of the instantaneous spectroscopic FOV of PACS (blue) and SAFARI (yellow). 
The large green box shows the area covered by SAFARI in a 900\,hour spectral full wavelength spectral 
survey ($\sim$1 degree).}
  \label{fig:millenium}
\end{figure}

\textbf{(ii) The formation and evolution of galaxies}: SPICA observations will provide a unique insight into the
basic questions about how galaxies form and evolve such as: What drives the evolution of the massive, dusty
distant galaxy population, and what feedback/interplay exists between the physical processes of mass accretion
and star formation? How and when do the normal, quiescent galaxies such as our own form, and how do they
relate to (Ultra) Luminous Infrared Galaxies (ULIRGs)? How do galaxy evolution, star formation rate and
AGN activity vary with environment and cosmological epoch?
Substantial progress in this area can only be made by making the transition from large-area photometric to
large-area spectroscopic surveys in the mid to far infrared, which will be possible with SPICA. This is because
the mid and far infrared region plays host to a unique suite of diagnostic lines to trace the accretion and star
formation, and to probe the physical and chemical conditions in different regimes from AGN to star-forming
regions. While the Herschel-PACS spectrometer detects the brightest far infrared objects at z$\sim$1, SPICA
will be able to carry out blind spectroscopic surveys out to z$\sim$3. This will lead to the first statistically 
unbiased determination of the co-evolution of star formation and mass accretion with cosmic time. 
Spectroscopic surveys will provide direct and unbiased information on the evolution of the large scale structure 
in the Universe from z$\sim$3 and the unprecedented possibility to investigate the impact of environment on 
galaxy formation and evolution as a function of redshift.
The high sensitivity of SPICA will enable photometric surveys beyond z$\sim$4 that will resolve more than
90\% of the Cosmic Infrared Background (in comparison with 50\% that Herschel will achieve). SPICA will also
observe Milky Way type galaxies in the far infrared out to z$\sim$1, where the cosmic star formation rate peaks.
For the first time, we will be able to piece together the story of the evolution of our own galaxy and answer the
question of whether we are in a ''special place'' in the cosmos.\\


By observing at far-IR wavelengths, Herschel is presently providing spectacular results in many fields of Astronomy.
A new mission reaching better sensitivity in the mid- and far-IR domain
must therefore follow to continue this work  and gain a deeper understanding of the physics 
of the objects discovered by Herschel: \textit{i.e.,}~to spectroscopically measure star formation rates in 
the obscured extra-galactic sources at different cosmic epochs and explore the physics and chemistry of planet formation. To obtain this increase in sensitivity we need
a cold ($<$6\,K) telescope. SPICA promises this in 2018-2020.

SPICA will be an international mission and will be open to the worldwide astronomical
community.
Japan is in charge of the whole integration of 
the system. 
The assessment study on the European contribution to the SPICA project has been 
carried out under the framework of the ESA Cosmic Vision 
(\textit{SPICA Assessment Study Report for ESA Cosmic Vision 2015-2025 Plan}, see 2010arXiv1001.0709S
for details). 
US and Korean participation are also being discussed.\\\\
\textit{Acknowledgments}: We warmly thank Kate Isaak, Bruce Swinyard, Ana Heras and the SPICA/SAFARI science
teams for fruitful discussions over the last years. 
J.R.G.~was supported by a
\textit{Ram\'on y Cajal} contract and by the AYA2009-07304 and CSD2009-00038 grants from
the Spanish MICINN.



\begin{thebibliography}{99}
\bibitem[2010]{pil10} Pilbratt, G. L. et al. 2010, \textit{A\&A},  518, L1.
\bibitem[2010]{spr06} Springel, V.,  Frenk, C.S. \& White, S.D.M. 2006, \textit{Nature}, 440, 1137.
\bibitem[2010]{spi10} SPICA Study Team Collaboration, 2010.
\textit{SPICA Assessment Study Report for ESA Cosmic Vision 2015-2025 Plan}, arXiv1001.0709S.
\bibitem[2010]{swi09} Swinyard, B., Nakagawa, T., et al. 2009, 
\textit{Experimental Astronomy}, 23, 1, 193-219.
\end{thebibliography}
\end{document}